\PassOptionsToPackage{final}{graphicx}
\documentclass[10pt,conference,table]{IEEEtran}
\IEEEoverridecommandlockouts%

\usepackage[backend=biber,bibencoding=utf8,style=ieee,natbib,mincitenames=1,maxcitenames=2,maxbibnames=99,autocite=inline,sortcites,labelnumber]{biblatex}
\DeclareFieldFormat*{citetitle}{\emph{#1}}

\usepackage[draft=false,pdfusetitle]{hyperref}
\hypersetup{
	pdfsubject={Software Engineering, Programming Languages},
	pdfkeywords={deep learning, refactoring, imperative programs, graph execution}}
\hypersetup{colorlinks=false,allcolors=blue,pdfborder={0 0 0},}

\usepackage[font=scriptsize,skip=0pt]{subcaption}

\usepackage[obeyDraft,colorinlistoftodos,prependcaption]{todonotes}

\usepackage[group-separator={,},group-minimum-digits={3},detect-all]{siunitx}
\DeclareSIUnit[number-unit-product = ]\percent{\char`\%}

\usepackage[export]{adjustbox}
\usepackage{nicefrac}
\usepackage{textcomp}
\usepackage{soul}
\usepackage{mathrsfs}
\usepackage{amsmath}
\usepackage{amsfonts}
\usepackage{mathtools}
\usepackage[capitalize]{cleveref}

\crefname{listing}{Listing}{Listings}
\Crefname{listing}{Listing}{Listings}
\crefname{sublisting}{Listing}{Listings}
\Crefname{sublisting}{Listing}{Listings}

\usepackage{comment}
\usepackage{ellipsis}

\usepackage{algorithm}
\usepackage{algorithmic}

\crefname{ALC@unique}{Line}{Lines}
\Crefname{ALC@unique}{Line}{Lines}

\crefname{paragraph}{Section}{Sections}
\Crefname{paragraph}{Section}{Sections}

\graphicspath{{./graphics/}} %

\usepackage[shortlabels,inline]{enumitem}

\usepackage[
    font=scriptsize,
    skip=1pt,
]{caption}
\crefname{section}{\S}{\S}
\crefformat{section}{\S#2#1#3}
\Crefname{section}{\S}{\S}
\Crefformat{section}{\S#2#1#3}
\crefname{paragraph}{\S}{\S}
\Crefname{paragraph}{\S}{\S}
\crefname{figure}{fig.}{fig.}
\Crefname{figure}{Fig.}{Fig.}
\crefname{table}{tab.}{tab.}
\Crefname{table}{Tab.}{Tab}
\crefname{definition}{def.}{def.}
\Crefname{definition}{Def.}{Def.}
\crefname{listing}{lst.}{lst.}
\Crefname{listing}{Lst.}{Lst.}
\crefname{sublisting}{lst.}{lst.}
\Crefname{sublisting}{Lst.}{Lst.}
\setlist{noitemsep,nosep}
\setlist{leftmargin=*}
\usepackage[compact]{titlesec}

\makeatletter
\newcommand\notsotiny{\@setfontsize\notsotiny\@vipt\@viipt}
\makeatother

\usepackage[
    draft=false,
    newfloat,
    finalizecache=false,
    frozencache=true
]{minted}
\newminted{diff}{
    numbersep=3pt,
    style=trac,
}
\newminted{python}{}
\newmintinline{python}{}
\definecolor{bg}{rgb}{0.95,0.95,0.95}
\newminted{pycon}{bgcolor=bg,linenos=false,tabsize=4}
\newmintinline{pycon}{}
\newminted{java}{}
\newmintinline{java}{}
\newmintinline{text}{
}
\setmintedinline{
    style=bw,
    fontsize=\small,
}
\setminted{%
    numbersep=2pt,
    style=trac,
    fontsize=\notsotiny,
    autogobble,
    breaklines,
    resetmargins,
    breakindentnchars=1,
    linenos,
    tabsize=1,
    breakafter=.,
    breakautoindent,
    breakaftersymbolpre=,
    escapeinside=||}

\newcommand{\pyi}[1]{\pythoninline{#1}}

\usepackage{booktabs}
\usepackage{tabulary}
\usepackage{tabularx}
\usepackage{threeparttable}
\usepackage{multirow}

\newlist{findings}{enumerate}{1}
\setlist[findings]{label=(\arabic*),ref=\arabic*}
\crefname{findingsi}{finding}{findings}
\Crefname{findingsi}{Finding}{Findings}

\newlist{bestpractices}{enumerate}{1}
\setlist[bestpractices]{label=(\arabic*),ref=\arabic*}
\crefname{bestpracticesi}{best practice}{best practices}
\Crefname{bestpracticesi}{Best practice}{Best practices}

\newlist{antipatterns}{enumerate}{1}
\setlist[antipatterns]{label=(\arabic*),ref=\arabic*}
\crefname{antipatternsi}{anti-pattern}{anti-patterns}
\Crefname{antipatternsi}{Anti-pattern}{Anti-patterns}

\newlist{recommendations}{enumerate}{1}
\setlist[recommendations]{label=(\arabic*),ref=\arabic*}
\crefname{recommendationsi}{recommendation}{recommendations}
\Crefname{recommendationsi}{Recommendation}{Recommendations}

\makeatletter
\def\TFF{\@ifstar\@TFF\@@TFF}
\def\@TFF{\pythoninline{tf.function}}
\def\@@TFF{\pythoninline{@tf.function}}
\makeatother

\newcommand{\eagtohyb}{\textsc{Convert Eager Function to Hybrid}}
\newcommand{\opthyb}{\textsc{Optimize Hybrid Function}}

\crefformat{preconditions}{P#2#1#3}

\begin{document}

\title{Towards Safe Automated Refactoring of Imperative Deep Learning Programs to Graph Execution\thanks{This material is based upon work supported by the National Science Foundation under Award Nos.~CCF-22-00343 and CNS-22-13763.}} %

\author{Raffi Khatchadourian\IEEEauthorrefmark{1}\IEEEauthorrefmark{2}, Tatiana Castro V\'{e}lez\IEEEauthorrefmark{2}, Mehdi Bagherzadeh\IEEEauthorrefmark{3}, Nan Jia\IEEEauthorrefmark{2}, Anita Raja\IEEEauthorrefmark{1}\IEEEauthorrefmark{2}\\
\IEEEauthorblockA{
	\IEEEauthorrefmark{1}City University of New York (CUNY) Hunter College,
	\IEEEauthorrefmark{2}CUNY Graduate Center,
	\IEEEauthorrefmark{3}Oakland University\\
	Email: raffi.khatchadourian@hunter.cuny.edu, tcastrovelez@gradcenter.cuny.edu, mbagherzadeh@oakland.edu,\\
	    njia@gradcenter.cuny.edu, anita.raja@hunter.cuny.edu}
}

\onecolumn%
\listoftodos%
\twocolumn%

\maketitle

\begin{abstract}

    Efficiency is essential to support responsiveness w.r.t.~ever-growing datasets, especially for Deep Learning (DL) systems. DL frameworks have traditionally embraced \emph{deferred} execution-style DL code---supporting symbolic, graph-based Deep Neural Network (DNN) computation. While scalable, such development is error-prone, non-intuitive, and difficult to debug. Consequently, more natural, imperative DL frameworks encouraging \emph{eager} execution have emerged at the expense of run-time performance. Though hybrid approaches aim for the ``best of both worlds,'' using them effectively requires subtle considerations to make code amenable to safe, accurate, and efficient graph execution.
We present our ongoing work on automated refactoring that assists developers in specifying whether and how their otherwise eagerly-executed imperative DL code could be reliably and efficiently executed as graphs
while preserving semantics. The approach, based on a novel imperative tensor analysis,
will automatically determine when it is safe and potentially advantageous to migrate imperative DL code to graph execution and modify decorator parameters or eagerly executing code already running as graphs. The approach is being implemented as a PyDev Eclipse IDE plug-in and uses the WALA Ariadne analysis framework. We discuss our ongoing work
towards optimizing imperative DL code to its full potential.

\end{abstract}


\section{Introduction}\label{sec:intro}

Machine Learning (ML), including Deep Learning (DL), systems are pervasive. They use dynamic models, whose behavior is ultimately defined by input data. However, as datasets grow, efficiency becomes essential~\cite{Zhou2020}.
DL frameworks have traditionally embraced a \emph{deferred} execution-style that supports symbolic, graph-based Deep Neural Network (DNN) computation~\cite{Google2021,Chen2015}. While scalable, development is error-prone, cumbersome, and produces programs that are difficult to debug~\cite{Zhang2018,Islam2019,Islam2019a,Zhang2019}.
Contrarily, more natural, less error-prone, and easier-to-debug \emph{imperative} DL frameworks~\cite{Agrawal2019,Paszke2019,Chollet2020} encouraging \emph{eager} execution have emerged. Though ubiquitous,
such programs are less efficient and scalable as their deferred-execution counterparts~\cite{Chen2015,Paszke2019,Moldovan2019,Facebook2019,Jeong2019,Google2022}.
Thus, hybrid approaches~\cite{Moldovan2019,Facebook2019,Apache2021b}
execute imperative DL programs as static graphs at run-time. For example, in \citetitle{Abadi2016}~\cite{Abadi2016},
\citetitle{Moldovan2019}~\cite{Moldovan2019} can enhance performance by decorating (annotating)---with optional yet influential decorator arguments---appropriate Python function(s) with \TFF. Decorating functions with such hybridization %
APIs
can increase
code performance without explicit modification. %

Though promising, hybridization
necessitates non-trivial
metadata~\cite{Jeong2019} and exhibits limitations and known issues~\cite{Google2021b} with native program constructs. Subtle considerations are required to make code amenable to safe, accurate, and efficient graph execution~\cite{CastroVelez2022,Cao2021}.
Alternative approaches~\cite{Jeong2019} impose custom Python interpreters, which may be impractical for industry, and support only specific Python constructs. Thus, developers are burdened with making their code compatible with the underlying execution model conversion and \emph{manually} specifying the functions to be converted.
Manual analysis and refactoring (semantics-preserving, source-to-source transformation)
can be overwhelming, error- and omission-prone~\cite{Dig2009}, and
complicated by
Object-Orientation (OO)
(e.g., \citetitle{Chollet2020}~\cite{Chollet2020})
and dynamically-typed languages (e.g., Python).

We present our ongoing work on a fully automated, semantics-preserving refactoring approach that transforms otherwise eagerly-executed imperative (Python) DL code for enhanced performance by specifying whether and how such code could be reliably and efficiently executed as graphs at run-time. The approach---based on a novel tensor analysis specifically for imperative DL code---will infer when it is safe and potentially advantageous to migrate imperative DL code to graph execution and modify decorator parameters or eagerly executing code already running as graphs.
It will also discover possible side-effects in
Python functions to safely transform imperative DL code to either execute eagerly or as a graph at run-time.
While LLMs~\cite{OpenAI2023} and
big data-driven refactorings~\cite{Dilhara2022}
have emerged, obtaining a (correct) dataset large enough to automatically extract the proposed refactorings is challenging as developers struggle with (manually) migrating DL code to graph execution~\cite{CastroVelez2022}. Also, while developers generally underuse automated refactorings~\cite{Negara2013,Kim2012}, since data scientists and engineers may not be classically trained software engineers, they may be more open to using automated (refactoring) tools. Furthermore, our approach will be fully automated with minimal barrier to entry. Our refactoring approach is being implemented as an open-source PyDev Eclipse Integrated Development Environment (IDE) plug-in~\cite{Zadrozny2023} that integrates analyses from the WALA Ariadne analysis framework~\cite{Dolby2018}. Moreover, while the refactorings will operate on imperative DL code that is easier-to-debug than its deferred-execution counterparts, the refactorings themselves will not improve debuggability but instead enable developers to have \emph{performant} easily-debuggable (imperative) DL code.

\section{Motivating Examples}\label{sec:motive}

\begin{listing}
    \begin{minipage}[t]{0.5\linewidth}
	\begin{pythoncode*}{numbersep=2pt}
		||
		class SequentialModel(Model):|\label{lne:subclass}|
			def __init__(self, **kwargs):
				super(SequentialModel, self)
					.__init__(...)
				self.flatten = layers.Flatten(
					input_shape=(28, 28))
				num_layers = 100 # Add layers.
				self.layers = [layers
					.Dense(64,activation="relu")
						for n in range(num_layers)]
				self.dropout = Dropout(0.2)
				self.dense_2 = layers.Dense(10)


			def __call__(self, x):
				x = self.flatten(x)
				for layer in self.layers:
					x = layer(x)
				x = self.dropout(x)
				x = self.dense_2(x)
				return x
	\end{pythoncode*}
	\subcaption{Code snippet before refactoring.\label{lst:model_before}}
    \end{minipage}\hfill
    \begin{minipage}[t]{0.5\linewidth}
	\begin{pythoncode*}{numbersep=2pt}
		|\smash{\ul{\mbox{import tensorflow as tf}}}\label{lne:tfimp}|
		class SequentialModel(Model):
			def __init__(self, **kwargs):
				super(SequentialModel, self)
					.__init__(...)
				self.flatten = layers.Flatten(
					input_shape=(28, 28))
				num_layers = 100 # Add layers.
				self.layers = [layers
					.Dense(64,activation="relu")
						for n in range(num_layers)]
				self.dropout = Dropout(0.2)
				self.dense_2 = layers.Dense(10)

			|\smash{\ul{@tf.function}}\label{lne:tfunc}|
			def __call__(self, x):
				x = self.flatten(x)
				for layer in self.layers:
					x = layer(x)
				x = self.dropout(x)
				x = self.dense_2(x)
				return x
	\end{pythoncode*}
	\subcaption{Improved code via refactoring.\label{lst:model_after}}
    \end{minipage}
    \caption{\citetitle{Abadi2016} imperative (OO) DL model code~\cite{Google2022}.\vspace{-1.25em}}\label{lst:model}
    \vspace{-1em}
\end{listing}

\Cref{lst:model_before} portrays \citetitle{Abadi2016} imperative (OO) DL code representing a modestly-sized model for classifying images. By default, this code runs eagerly; however, it may be possible to enhance performance by executing it as a graph at run-time. \Cref{lst:model_after}, lines~\ref{lne:tfimp} and \ref{lne:tfunc} %
display the
refactoring with the imperative DL code executed as a graph at run-time (added code is \ul{underlined}).
\citetitle{Moldovan2019}~\cite{Moldovan2019} is now used to potentially improve performance by decorating---with optional yet influential decorator arguments---\pyi{call()}
with \TFF. %
At run-time, \pyi{call()}'s execution will be ``traced'' and an equivalent graph will be generated~\cite{Google2021b}. In this case, a speedup ($\nicefrac{\mathit{run time}_{\mathit{old}}}{\mathit{run time}_{\mathit{new}}}$) of $\sim$\num{9.22}
ensues~\cite{Khatchadourian2021}. %
Though promising, using hybridization reliably \emph{and} efficiently is challenging~\cite{Jeong2019,Google2021b}.
For instance, side-effect producing, native Python statements
are problematic for \TFF*-decorated functions~\cite{Google2021b}.
Because their executions are traced, a function's behavior is ``etched'' (frozen) into its corresponding graph and thus can have unexpected results.

\section{Optimization Approach}\label{sec:approach}

We work towards two new refactorings, namely, \eagtohyb\ and \opthyb.\ The former transforms otherwise eagerly-executed imperative (Python) DL code for enhanced performance, automatically specifying whether and how such code could be reliably and efficiently executed as graphs at run-time. It infers when it is \emph{safe} and potentially \emph{advantageous} to migrate imperative DL code to graph execution. The latter either modifies existing decorator parameters or the structure of imperative DL code \emph{already} running as graphs.
While the DL code portrayed in \cref{lst:model_after} is sequentially executed, hybrid functions share some commonality with concurrent programs. For example, to avoid unexpected behavior, such functions should avoid side-effects. In our refactoring formulation, we will approximate aspects like side-effects in deciding which transformations to perform to ensure that they are safe, i.e., that the original program semantics are preserved. To ensure that the transformations are advantageous, we will involve (imperative) tensor analysis to avoid function ``retracing'' so that newly hybridized functions have tensor parameters whose shapes are sufficiently general. Otherwise, the transformed function would be traced \emph{every} time it called, potentially \emph{degrading} performance~\cite{Google2021p}. Furthermore, DL code interacts with many third-party libraries~\cite{Dilhara2021,Islam2019,Islam2020a,Islam2019a,Zhang2021}. Our approach will operate on a \emph{closed-world} assumption that assumes access to all source code that could possibly affect or be affected by the refactorings. We will then relax this assumption in our implementation by conservatively failing refactoring preconditions for functions defined elsewhere.

Challenges include a lack of static type information, which is necessary to determine candidate functions (must have at least one parameter of type \pyi{Tensor}). Our current approach is to use Python 3 type hints if present. We are also augmenting \citetitle{Dolby2018}~\cite{Dolby2018} to analyze imperative Python code (TensorFlow 2). Also, unlike, e.g., Java, Python has no restrictions on decorator (annotation) arguments. Thus, we utilize \citetitle{Dolby2018} for dataflow analysis to determine configuration values. Furthermore, \TFF*\ may be used as a first-class function instead of a decorator. To this end, we are working towards building a fluent API typestate analysis for imperative DL code by adapting the work of \citet{Khatchadourian2019}.
Existing work for determining tensor shapes
only works for \emph{procedural} TensorFlow (TF v1) code. Even finding a fully qualified name (FQN) of a program entity (e.g., \pyi{tf.function}) statically is difficult in Python. Import statements can appear anywhere in the code. Moreover, there is also import aliasing (e.g., \pyi{import tensorflow as tf}) to handle.

Complex static analysis can be expensive and not scalable. However, such analyses may be useful for future approaches by the community. Faster \emph{speculative} analysis~\cite{Zhou2020} uses contextual (ML) keywords to make assumptions. Here, assumptions are explicitly presented to developers. Developers then decide if assumptions are valid and may \emph{reject} the refactoring. However, it involves more developer input, and DL frameworks evolve constantly, potentially changing ``keywords.'' We are currently considering devising a hybrid analysis~\cite{Jeong2019} that runs DL code for several epochs to collect type information. Such an approach can be fast but relies on particular datasets and thus may be less generalizable.

\section{Conclusion \& Future Work}\label{sec:conc}

Imperative DL code is easier to debug, write, and maintain than traditional DL code that runs in a deferred execution. However, it comes at the expense of (run-time) performance. Hybrid approaches bridge the gap between eager and graph execution. Using hybrid techniques to achieve optimal performance and semantics preservation is difficult. Our in progress work aims to automate client-side analyses and transformations to \emph{use} hybridization APIs correctly and optimally. In the future, we plan to evaluate our approach by curating a DL Python project dataset from~\citet{CastroVelez2022} that is ready to be analyzed, with dependency and build infrastructure. We will also open-source our refactoring tools.

\printbibliography%

\end{document}